\documentclass[useAMS,usenatbib,usegraphicx]{mn2e}

\title[Spectral and polarization study of the double radio relics in
A3376]{Spectral and polarization study of the double relics in Abell 3376 using
the GMRT and the VLA}
\author[Ruta Kale, K. S. Dwarakanath, Joydeep Bagchi, Surajit Paul]{Ruta
Kale$^{1,2,3,4}$\thanks{E-mail:
rkale@ira.inaf.it}, K. S. Dwarakanath$^{4},$ 
Joydeep Bagchi$^{3}$, and Surajit Paul$^{3}$\\
$^{1}$Dipartimento di Astronomia, Universita di Bologna, via Ranzani 1, I-40127 Bologna, Italy\\
$^{2}$INAF-Istituto di Radioastronomia, Via Gobetti 101, I-40129 Bologna, Italy\\
$^{3}$Inter University Centre for Astronomy and Astrophysics, Post Bag 4,
Ganeshkhind, Pune 
University Campus, Pune 411007, India\\
$^{4}$Raman Research Institute, C. V. Raman Avenue, Sadashivanagar, Bangalore
560080, India}
\begin{document}

\date{accepted in MNRAS}

\pagerange{\pageref{firstpage}--\pageref{lastpage}} \pubyear{2011}

\maketitle

\label{firstpage}

\begin{abstract}
Double radio relics in galaxy clusters are rare phenomena that trace shocks in
the outskirts of merging galaxy clusters. We have carried out a spectral and
polarization study of 
the spectacular double relics in the galaxy cluster A3376 using the Giant
Metrewave Radio 
Telescope at 150 and 325 MHz and the Very Large Array at 1400 MHz.
The polarization study at 1400 MHz reveals a high degree of polarization
($\sim30\%$) and aligned magnetic 
field vectors (not corrected for Faraday rotation) in the eastern relic.
 A highly polarized ($>60\%$)
filamentary radio source of size $\sim300$ kpc near the 
eastern relic and north of the bent-jet radio galaxy is detected for the
first time. The western relic is less polarized and does not show aligned
magnetic field vectors. 
The distribution of spectral indices between 325 and 1400 MHz 
over the radio relics show steepening from the outer to the inner edges of the
relics. The spectral indices of the eastern and the western relics 
imply Mach numbers in the range 2.2 to 3.3.
Remarkable features such as the inward filament extending from the
E-relic, the highly polarized filament, the complex polarization
properties of the W-relic and the
separation of the BCG from the ICM by a distance $>900$ kpc are noticed in the
cluster.
A comparison with simulated cluster mergers
is required to understand the complex properties of the double relics in the
context of the merger in A3376. An upper limit (log(P$_{1.4GHz}$ W
Hz$^{-1}$)$<23.0$) on
the strength of a Mpc size radio halo in A3376 is estimated.
\end{abstract}


\section{Introduction}
Clusters of galaxies, being the largest gravitationally bound structures 
in the Universe are interesting sites to study extreme phenomena. Evolution of
galaxy clusters occurs through mergers of galaxy groups and galaxy sub-clusters
involving dissipation of gravitational binding energies $\sim10^{64}$ erg 
in the intra-cluster medium (ICM) via shocks and turbulence \citep{ryu03}.
There is growing evidence that
cluster-cluster mergers are connected to the presence of diffuse large scale
radio sources 
called radio halos and radio relics \citep{fer08}. 
The channeling of the gravitational energy into the
acceleration of particles and possibly amplification of the cluster magnetic 
field are not well understood and are currently the topics of active research.

Radio halos are $\sim$Mpc size radio sources, located co-spatially with the
X-ray emission from the ICM.
Acceleration of particles in the ICM through turbulence and production of 
relativistic electrons as secondary products of hadronic collisions in the 
ICM are the two mechanisms proposed for the generation of radio
halos \citep[see][and references therein]{bru11}. 

Radio relics are elongated, polarized, filamentary and sometimes arc-like
radio sources. One class of radio relics has been proposed to be
remnants of radio galaxies evolving passively in the surrounding medium or
revived by adiabatic
compression by environmental shock waves \citep{ens01}. The solitary
filamentary relics of sizes $\sim50-500$ kpc, sometimes located close to
radio galaxies in clusters are likely to have these origins
\citep{kal11}. The radio relics of the other class have arc-like
morphologies, $\sim$Mpc extents and are located at the peripheries of galaxy
clusters. These are believed to be tracers of the cluster merger shocks or
accretion shocks
\citep{ens98}. The shocks are stronger toward the outskirts 
of clusters where the density of electrons is very low (n$_e<10^{-4}$
cm$^{-3}$).
Therefore, detecting them in X-rays is a challenge \citep{aka11}. The radio
relics thus offer a unique window to trace the shocks and the ICM at cluster
peripheries.

Shocks are known to accelerate electrons by the diffusive shock
acceleration (DSA) mechanism and compress magnetic fields along the direction of
 propagation \citep{dru83}. The DSA produces a power law
distribution
of relativistic electrons and thus provides a link between the spectral
index of the radio emission and the strength of the shock. The alignment of
magnetic 
fields results in polarized radio emission from radio relics. Thus, the
integrated spectra, spectral index distribution and polarization 
of radio relics are crucial to find the properties 
of shocks in the ICM. However, radio relics are rare and only a handful of them
are known. Therefore studying the properties of the known sources in as much
detail as possible is important.
Even rarer are the cases when the arc-like relics occur in pairs
around galaxy clusters. Such relics occur in
cluster-cluster mergers with the merger axes nearly in
the plane of the sky. The effects of projection in studying the shocks and
the merger in such a configuration are expected to be minimal.
 There are only nine clusters where such ``double relics''
have been discovered, namely, ZwCl $2341.1+0000$ \citep{bag02,wee09}, 
A3376, PLCK $G287.0+32.9$ \citep{bag02,bag06,bag11}, A3667 \citep{rot97}, ZwCl
$0008.8+5215$,
CIZA J$2242.8+5301$ \citep{wee11c,wee10}, RXCJ $1314.4-2515$ \citep{fer05}, 
A2345 \citep{gio99,bon09} and A1240 \citep{kem01,bon09}. There are three more
candidate
double radio relics, namely $0217+70$ \citep{bro11}, A3365 \citep{wee11b} and 
A1758N \citep{gio09}.

In this paper we present a polarization and spectral study of the double relics
in A3376 using the Giant
Metrewave Radio Telescope (GMRT) and the Very Large Array (VLA).
Abell 3376 is a nearby ($z=$0.046) X-ray luminous ($L_X\sim 2\times10^{44}
h^{-2}_{70}$ erg s$^{-1}$) cluster with spectacular double relics
\citep{bag06}. It has an average temperature of $\sim 4$ keV and
the X-ray surface brightness has a comet-like appearance which indicates the
ongoing dynamical activity in the cluster. 
The virial mass of this cluster is estimated to be
$\sim3.64\times10^{14}h_{70}^{-1}$ M$_\odot$ \citep{gir98}. 
The two radio relics in A3376 are separated by a distance of $\sim$ 2 Mpc and 
are located 
almost symmetrically around the X-ray emission from the ICM. The previous study
of
these relics at 1.4 GHz revealed the basic morphology of these relics and
stated the possibilities that either outgoing cluster merger shocks or accretion
shocks are responsible for the relics
\citep{bag06}. In this paper we provide a more detailed description.

The paper is organised as follows. The observations and data
reduction are described in section 2. 
The properties of the relics revealed from the radio images and the spectral
index images are presented in section 3. In section 4 the implications of these
properties are discussed. The work is summarized in section 5.

We adopt a $\Lambda$CDM cosmology with $H_0$ = 71 km s$^{-1}$ 
Mpc$^{-1}$, $\Omega_M$ = 0.27, and 
$\Omega _\Lambda = 0.73$, resulting in a scale of 0.89 kpc arcsec$^{-1}$ for the
redshift, 0.046 of A3376. The spectral index, $\alpha$ is defined as
$S\propto\nu^\alpha$, 
where $S$ is the flux density and $\nu$, the frequency.
\section{Observations and Data Reduction}
The cluster A3376 was observed at 325 and 150 MHz with the GMRT in Nov. 2009
(proposal code $17\_024$, PI R. Kale). 
Two observing sessions of $\sim7$ hr duration centered on the
positions (J2000) RA 06h02m15s DEC -39d57m00s and RA 06h00m07s DEC
-40d02m00s were conducted at 325 MHz. A bandwidth of 32 MHz was used.
At 150 MHz, two observing sessions of $\sim 7$ hr duration with 
a bandwidth of 16 MHz were conducted. However data from one of the 150
MHz sessions were heavily corrupted by radio frequency interference (RFI) and
had to be discarded.

The data at 150 and 325 MHz were analyzed using the standard
procedures in NRAO
Astronomical Image Processing System (AIPS) version 31DEC10. The VLA
calibrator values (Perley-Butler 2010) were used
for absolute flux calibration. Wide field imaging was carried out to account for
the
non-coplanarity of the array. The sensitivities at both 325 and 150 MHz were 
improved by several iterations of self-calibration. A
single image was produced by combining 
images obtained from the data in the two observing sessions at 325 MHz. An 
image using the `natural' weighting scheme (robust$=5$ in AIPS) with a rms 2.0 
mJy beam$^{-1}$ and a
synthesized beam of $39''\times39''$ was produced at 325 MHz
(Fig.~\ref{pband}). 
Similarly, an image with a rms $5.0$ mJy beam$^{-1}$ and a synthesized beam of  
$47''\times47''$ was produced at 150 MHz (Fig.~\ref{low}). 
High resolution images using the `uniform' weighting scheme 
(robust$=-5$ in AIPS) were produced at both, 150 and 325 MHz to identify  
discrete radio sources.

Archival data from the VLA (proposal code AB1057, PI J. Bagchi) at 1400 MHz in 
the configurations CnB and DnC were analyzed. It consisted of separate
pointings 
for each of the relics in the two configurations. 
Standard procedures of data editing, 
phase and flux calibration were carried out in AIPS (version 31DEC11). The 
 calibrator source $0616-349$ was used to obtain the polarization 
leakage terms and the source 3C286 was used to calibrate the polarization 
position angle. After calibration, the data in the two configurations for 
each of the relic were combined separately using the task `DBCON'. These 
visibilities were used to make stokes I, Q, U and V images. Self calibration 
was 
carried out to improve the sensitivities. The polarized intensity images were
corrected for the Ricean bias using the task `POLCO'. Stokes I
images 
(rms
$\sim40 \mu$Jy beam$^{-1}$) and Q and U images (rms $\sim30 \mu$Jy beam$^{-1}$)
were 
obtained (Fig.~\ref{pol}). The polarization vectors are not corrected for
Faraday rotation.

The construction of spectral index distribution maps requires radio images at  
two frequencies with closely matched
$uv$-plane sampling. Often, it is not possible to meet this requirement with a
single radio telescope. 
The combination of the GMRT at 325 MHz and the VLA in CnB and DnC 
configurations at 1.4 GHz provides matched uv-coverages, sufficient to obtain 
spectral index 
distributions over the double relics in A3376.
We obtained images with the
$uv$-distances limited to 0.125-25 k$\lambda$ and the weighting  
scheme of robust$=-5$ (pure uniform) and $uv-$taper of 2k$\lambda$ 
at both 325 and 1400 MHz. The uniform weights ensure same $uv$-plane weighting
distribution at the two frequencies. Primary beam gain corrections were
applied and the regions with flux densities below $5\sigma$ were blanked. These
images,
convolved to a common resolution of $69''\times69''$ were
used to construct the spectral index images and the corresponding uncertainty 
images (Fig.~\ref{spix}).
\section{Results: Properties of the double relics}
\begin{figure*}
\includegraphics[width=140 mm]{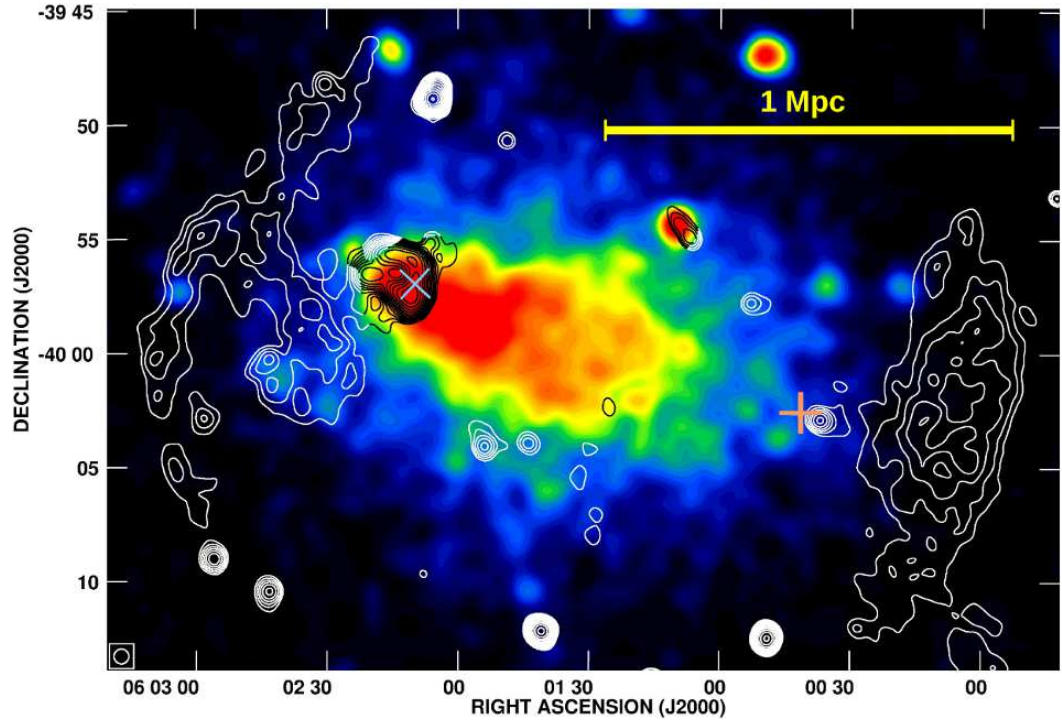}
\caption{The GMRT 325 MHz
image is shown in contours. The contour levels start 5.8
 mJy beam$^{-1}$ and increase by a factor of $\sqrt{2}$.
The synthesized beam is $39''\times39''$. The ROSAT soft X-ray band
image (0.1-2.4 keV) is shown in color. The $'\times'$ and the $'+'$ signs mark
the positions of the galaxies MRC $0600-399$ (bent double radio galaxy) and ESO
$307-13$ (BCG), respectively.}
\label{pband}
\end{figure*}

The cluster A3376 consists of two arc-like relics located almost
symmetrically 
around the X-ray emission from the ICM (Fig.~\ref{pband}). We will
henceforth refer to the relic to the east of the X-ray emission as E-relic and
the one 
to the west as the W-relic.

\begin{figure*}
\includegraphics[width=100mm, angle=-90]{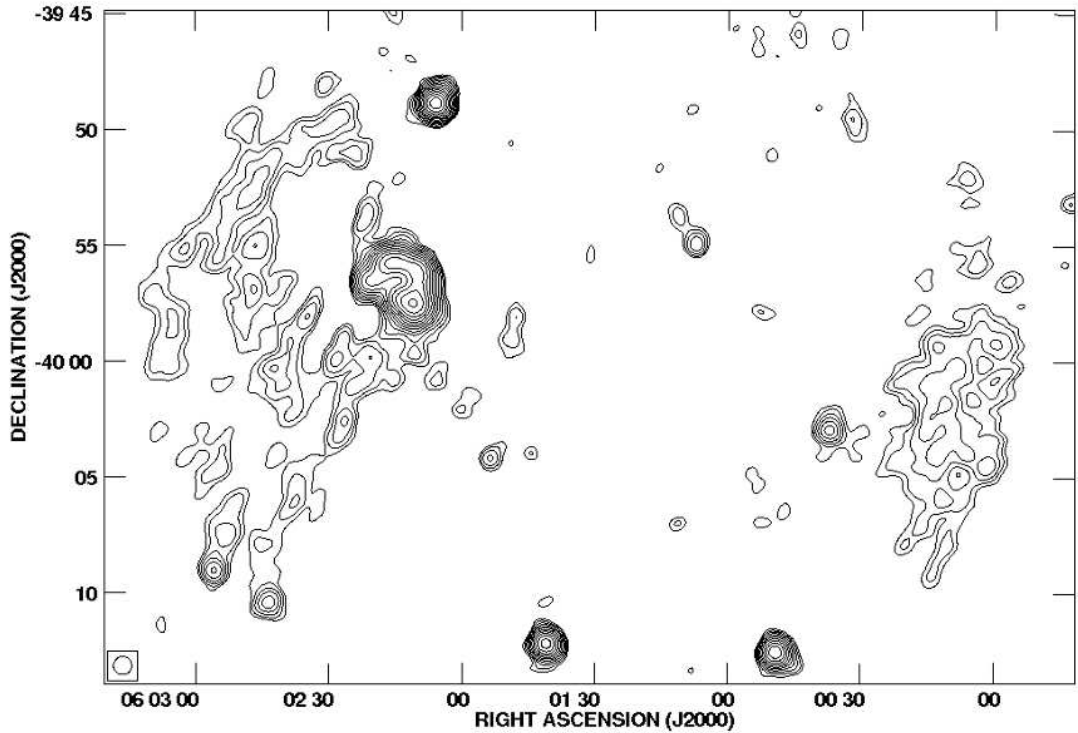}
\caption{The GMRT 150 MHz image is shown in contours. 
The contour levels start at 15.0
mJy beam$^{-1}$ and increase by a factor of $\sqrt{2}$.
The synthesized beam is $47''\times47''$.}
\label{low}
\end{figure*}

\begin{figure*}
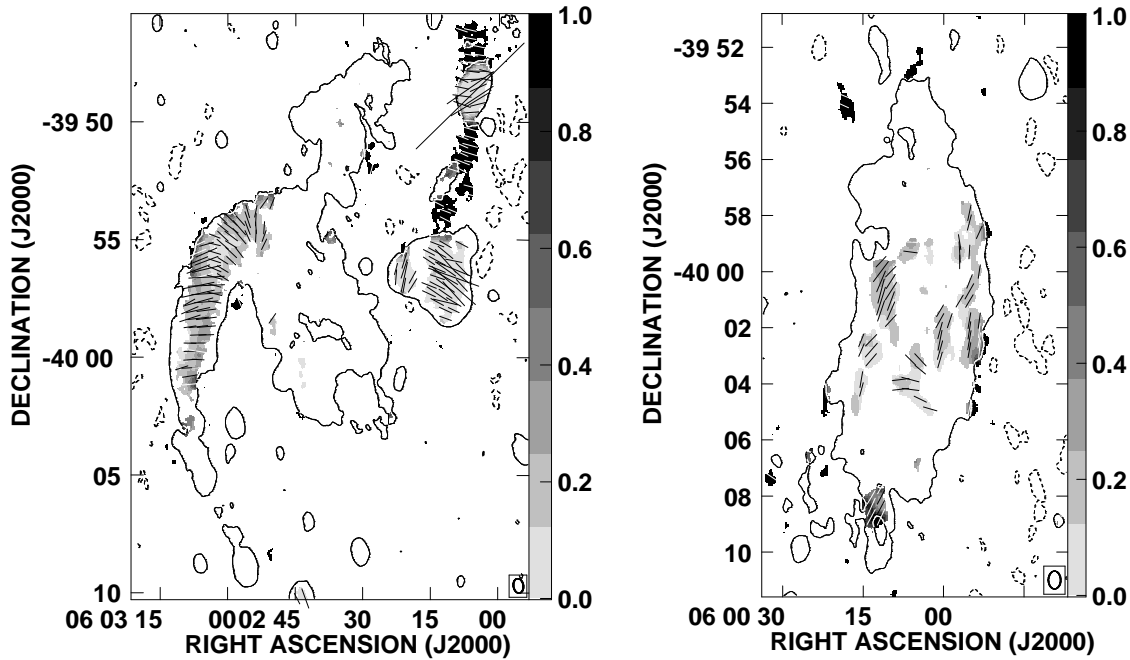

\includegraphics[height=83mm,angle=-90]{L_POL_E.eps}
\includegraphics[height=70mm,angle=-90]{L_POL_W.eps}
\caption{VLA 1400 MHz: Electric field vectors overlaid on fractional polarized
intensity shown
in grey-scale and Stokes I contours shown at -0.16, 0.16 mJy beam$^{-1}$. The
length of the vectors is proportional to the polarized intensity.
The synthesized beam is $37''\times25''$ (P. A. 9.22$^{\circ}$) in
 the left panel and $38''\times26''$ (P. A. 0.17$^\circ$) in the right
panel.}
\label{pol}
\end{figure*}

\begin{figure*}
\includegraphics[height=100mm]{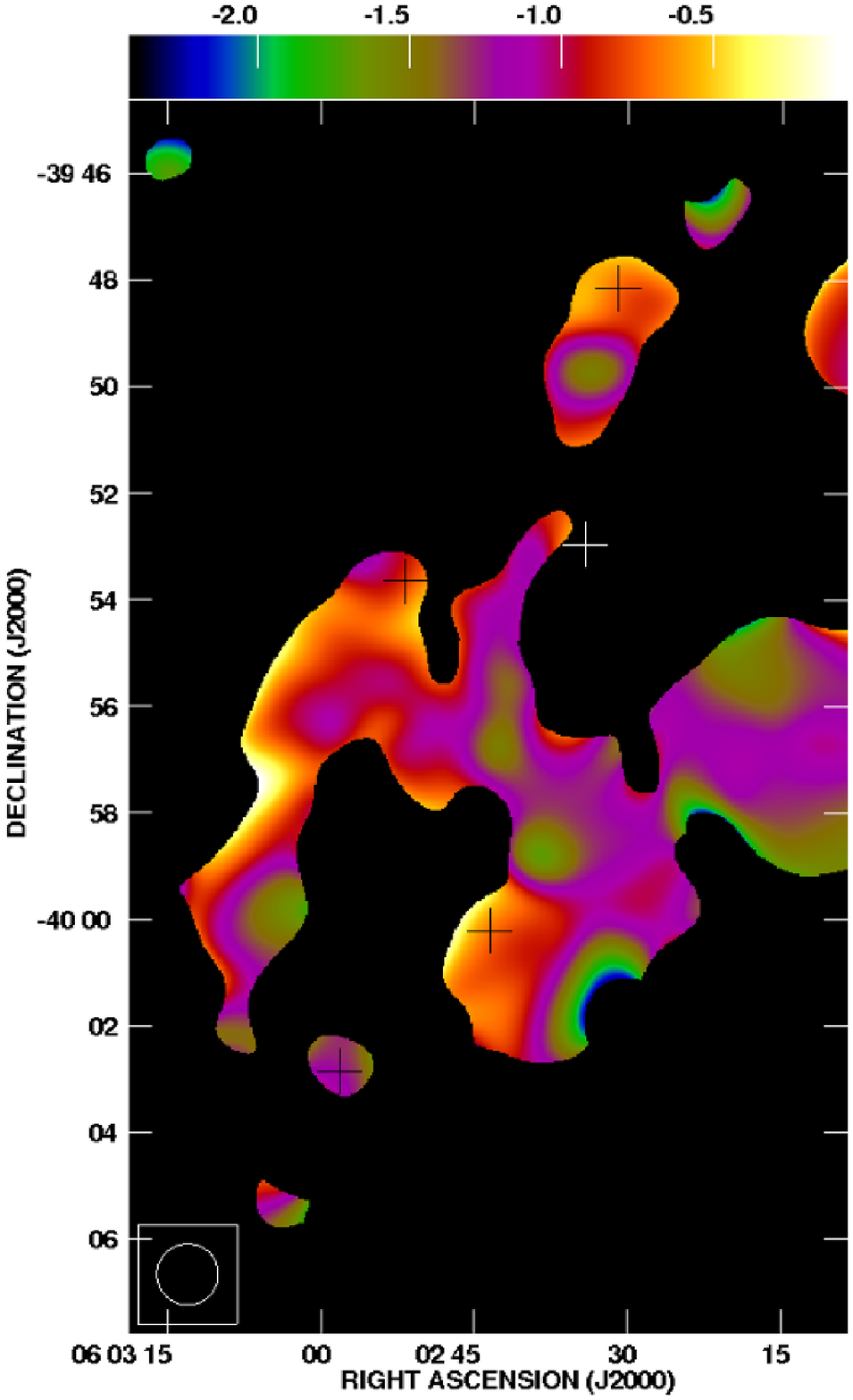}
\includegraphics[height=100mm]{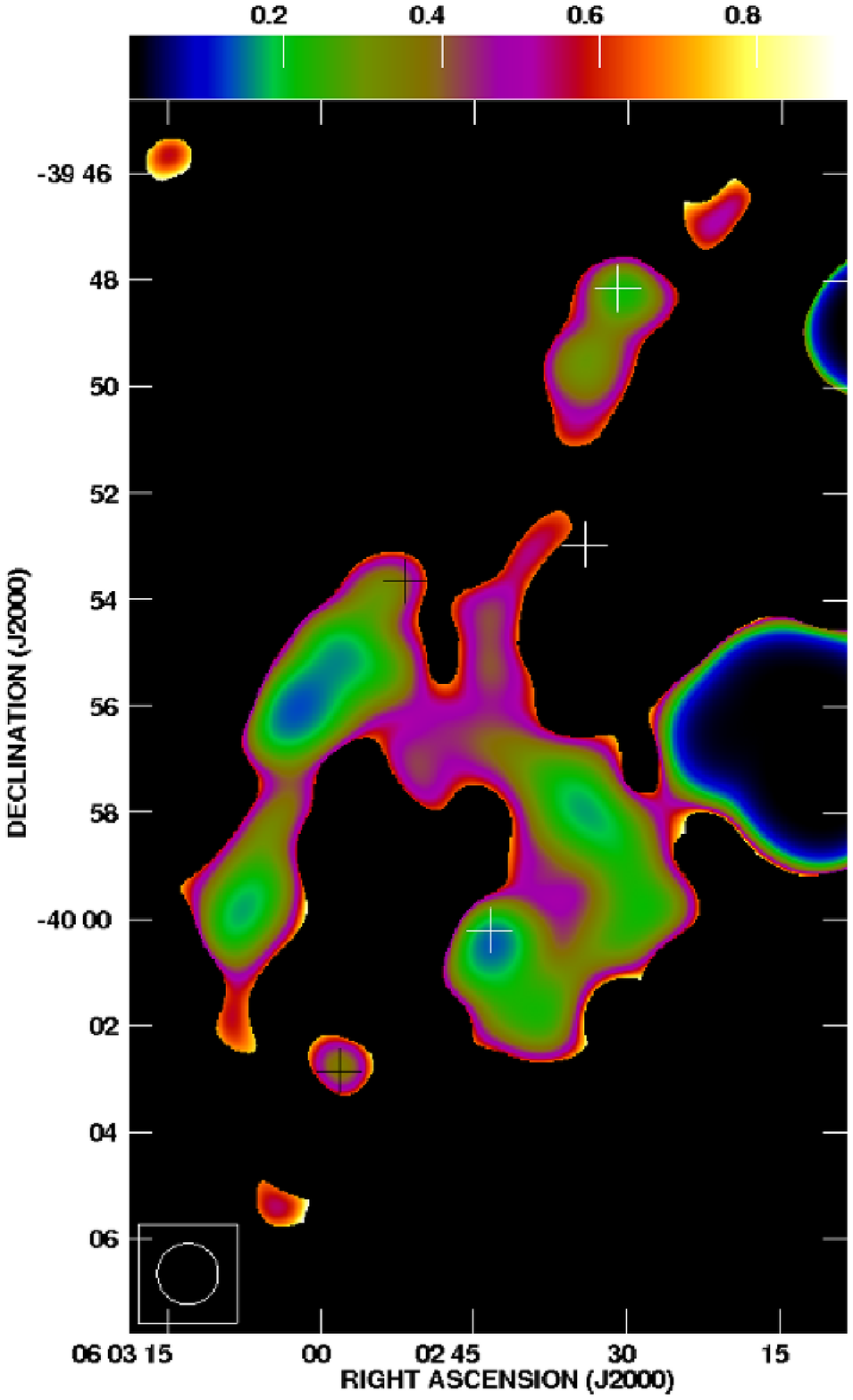}\\
\vspace*{0.8 cm}
\includegraphics[height=100mm]{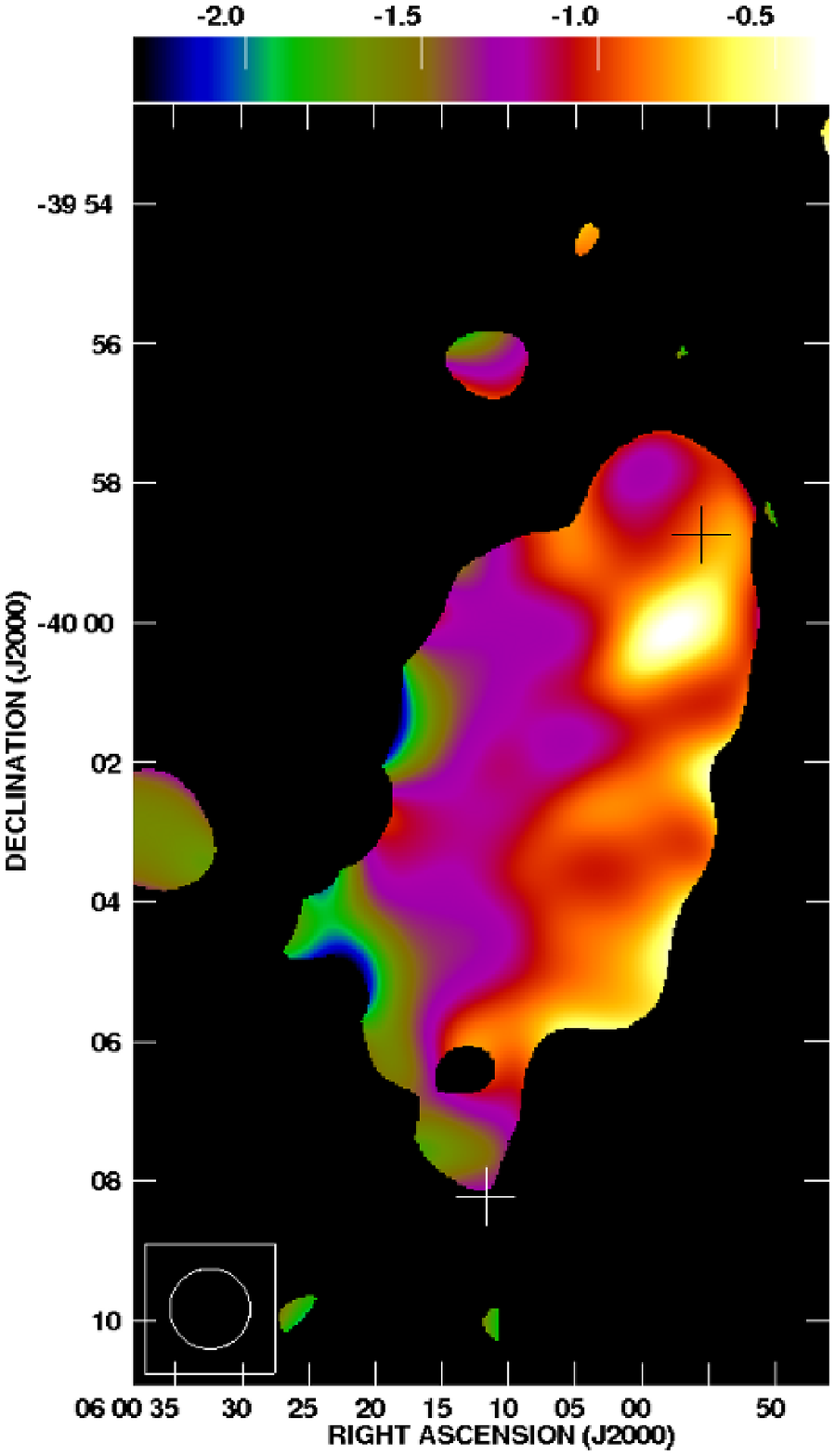}
\includegraphics[height=100mm]{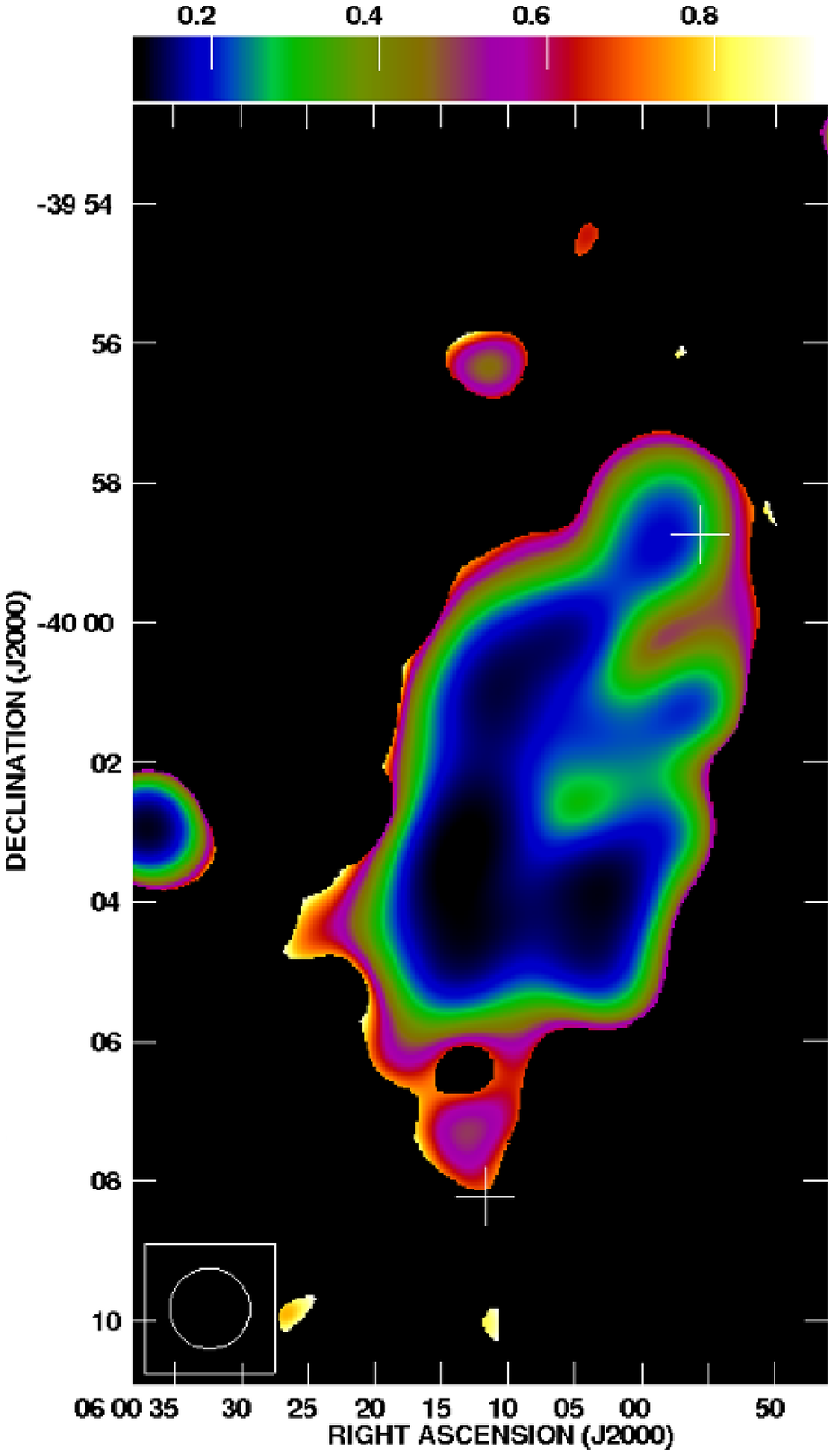}
\caption{The top two panels show the spectral index map between 325 and 1400 MHz
(left) and the 
corresponding uncertainty map (right) for the E-relic in colour. The bottom two
panels
show the same for the W-relic. 
The synthesized beam is $69''\times69''$ in
all the panels. The crosses mark the positions of peaks of the compact sources
in the regions of the relics, detected in the 1400 MHz images.}
\label{spix}
\end{figure*}

The E-relic can be divided into three regions- an arc-like southern
region, a relatively faint northern region and a 
filament extending into the cluster (Fig.~\ref{pband}). The southern arc has a 
sharp outer edge and a width in the range $\sim50-200$ kpc. A
fractional polarized intensity of 
up to $30\%$ is detected in it and the magnetic field vectors (uncorrected for
Faraday rotation) are aligned 
parallel to the elongation of the arc (Fig.~\ref{pol}, left). A steep 
gradient of spectral indices (-0.7 to -1.3) from the outer to the inner edge is
seen (Fig.~\ref{spix}, top left).  The northern portion
of the E-relic has steeper spectral indices (-1 to -1.5) and does not show a
sharp edge. There is a striking inward extending filament from the E-relic,
creating a `notch'  in the overall arc-like shape of the E-relic.
It has a linear extent of $\sim500$ kpc and a
steep spectral index ($\sim$-1.5). In the higher resolution ($20''\times20''$)
image at 1.4 GHz, the southern arc is prominent and the northern region together
with part of the inward filament forms another arc \citep[see Fig. 1B
in][]{bag06}. The inward filament and this arc connect to each other forming a
loop. These details are not resolved in the 325 MHz image presented here. 
It is clear that the E-relic, though apparently arc-like, has many other
complex structures associated with it, which require further study to
understand their real nature. The differences in spectral indices of the various
regions can also lead to differences in the morphology at widely separated
frequencies such as 150 - 1400 MHz. The regions of the E-relic having steeper 
spectra are detected in the 150 MHz image with higher significance than the
 southern arc prominent in 325 and 1400 MHz images (Fig.~\ref{low}).

The polarized intensity map of the E-relic also reveals a filament showing high 
fractional polarization ($>60\%$) located north of 
the bent-jet radio galaxy (Fig.~\ref{pol}, left).
The magnetic field vectors in this filament are also aligned approximately
parallel to those in the southern arc of the E-relic.

The outer edge of the W-relic also follows the curvature of the E-relic like a 
mirror image. The W-relic has a largest extent of $\sim900$ kpc and has a
maximum width of $\sim300$ kpc near the center (Fig.~\ref{pband}).
It is also polarized between $5$ to $20\%$ in a few pockets
but does not show alignment of magnetic field vectors in any single direction 
(Fig.~\ref{pol}, right). This may be an effect of Faraday rotation by the
magnetized ICM.
The distribution of spectral indices shows flatter
spectra toward the outer edge as compared to the inner edge (Fig.~\ref{spix},
bottom left). Arc-like radio relics typically show
monotonically decreasing surface brightness from their outer to inner edges.
However, the surface 
brightness of the W-relic shows a depression by a factor of $\sim 2$ at its
center. A larger error ($\sim0.3$) 
corresponding to this lower surface brightness region is seen in the spectral 
index uncertainty image (Fig.~\ref{spix}, bottom right). The W-relic at 150 MHz
lacks the arc-like 
outer edge that is seen at both 325 and 1400 MHz (Fig.\ref{low}).

The integrated flux densities and spectral indices of the relics are summarized
in Table ~\ref{fluxdens2}. The flux densities at 325 MHz and 150 MHz were
obtained from the images in Fig.~\ref{pband} and Fig.~\ref{low} respectively.
The images at 1400 MHz, using robust$=5$ (not shown), were used. The
contribution of unresolved sources estimated from high resolution images were
subtracted to obtain the estimates of the flux densities of the relics.
Both the relics show high frequency
steepening in their spectra. 
The largest detectable extent with the VLA at 1400 MHz ($\sim 18'$) is close to
the largest extents of the relics in the $\sim$north-south direction. There may
be some loss of flux density at 1400 MHz due to this resulting in an apparently
steep spectral index between 325- 1400 MHz.

\begin{table}
 \caption{Flux densities and spectral indices of the A3376 relics.}
 \label{fluxdens2}
 \begin{tabular}{lrr}
  \hline
  & E-relic &   W-relic  \\      
  \hline
$S_{1400}$ (mJy) & $122\pm10$& $113\pm10$ \\
$S_{325}$ (mJy) & $1770\pm90$& $1367\pm70$ \\
$S_{150}$ (mJy) &$3500\pm350$& $2962\pm300$\\
$\alpha_{150}^{325}$ &$-0.88\pm0.14$&$-1.00\pm0.14$\\
$\alpha_{325}^{1400}$ &$-1.82\pm0.06$&$-1.70\pm0.06$\\
  \hline
 \end{tabular}
\end{table}
\section{Discussion}
The cluster A3376 is a spectacular massive, merging cluster of galaxies. The
galaxy 
distribution shows atleast two sub-clusters located along the 
axis of elongation of the X-ray surface brightness \citep{fli06}. The giant
ring-like radio 
relics at the outskirts of the cluster are located along the same axis.
The alignment of magnetic field vectors in the E-relic and the spectral index 
steepening toward inner edges of the relics support the scenario of shock 
acceleration at the radio relics.

The acceleration of electrons at the shock by DSA results in a power law
spectrum of the electrons.
In a fully ionized plasma the slope of the power law
spectrum, $\delta_{inj}$, is related to the Mach
number, $M$, of the shock as given by \citep[e.g.][]{bla87}, 
\begin{equation}
 \delta_{inj} = -2 \frac{M^2 + 1}{M^2 - 1}
\end{equation}
The spectral index of the accelerated electrons is $\alpha_{inj} = (\delta_{inj}
+ 1) / 2$. 
If we assume that the shock is located at the radio relics and that the
synchrotron and inverse Compton losses have not affected the spectrum of the
accelerated electrons, then the flattest spectrum will
represent the spectral index at injection. The flattest spectral index in the
southern arc in the E-relic as read from the
spectral index map is $-0.70\pm0.15$ (Fig.~\ref{spix}, top left).
The implied Mach number is $3.31\pm0.29$. The spectral index at the outer edge
of the W-relic is $-1.0\pm0.2$ and the implied Mach number is
$2.23\pm0.40$ (Fig.~\ref{spix}, bottom left). These Mach numbers imply that the
shocks at the relics are cluster merger shocks and not accretion shocks
(M$>$10).

Deep X-ray observations with sensitive instruments can provide an independent
measure of the properties of shocks through the measurements of temperature
and pressure jumps at the location of radio relics.
 The $SUZAKU$ X-ray observations of A3376 by \citet{aka11b}
reveal temperature and
pressure jumps across the W-relic. A temperature jump from
$1.35\pm0.35$ keV in the pre-shock region to $4.81\pm0.29$ keV in the
post-shock region is reported at the W-relic.
They find Mach numbers of $2.94\pm0.60$ and $4.78\pm0.49$ from the
temperature and pressure jumps, respectively. The radio Mach number of the
W-relic is consistent with the one calculated using the temperature jump. 
The Mach number estimated from the pressure jump is less reliable due to
 the assumptions, such as of spherical symmetry, that are used in the
determination of the electron density profile \citep{aka11b}.

In the process of structure formation, large scale shocks are driven into the
diffuse medium. High Mach number shocks ($M>10$) are formed in filaments
surrounding
the virializing cores of clusters. On the other hand, low Mach number ($M<10$)
shocks are formed in the clusters. Most of the kinetic energy of the accreted
matter is dissipated by the relatively low Mach number shocks ($M<4$)
\citep{min00,ryu03}. These Mach numbers are consistent with those found
from the spectral indices of the A3376 radio relics and also other relics
\citep{wee10,fin10}. These findings imply shock velocities $\sim1000$ km
s$^{-1}$ at the radio relics. This shock velocity is comparable to that in
supernova remnants, however the Mach numbers are much lower \citep{jon12}. 
The low Mach number shocks are believed to be inefficient in accelerating
electrons 
from the thermal ICM to relativistic energies as required to produce the radio
relics \citep{kan02}. Thus the understanding of the generation of relativistic
electrons through DSA at such low Mach number shocks is still incomplete.
Nevertheless the coincidence of such shocks with the radio relics does imply
their role in the reacceleration of a pre-existing population of relativistic
electrons or in strengthening the magnetic fields.
\subsection{Remarkable features of A3376}
The spectral and polarization study of double radio relics in A3376 has
revealed several new features. The properties of the A3376 
relics such as -- aligned polarization vectors and spectral steepening from
outer to inner edges, are similar to the radio relics in other galaxy clusters.
Majority of the single and double relics at cluster outskirts have been
found to be located along the direction of elongation of the X-ray surface
brightness \citep{wee11b}. The A3376 relics are in line with this property 
 of other relics as well.

However there are some remarkable properties of the relics in A3376 and of the
cluster itself that we have noticed. 
The $\sim$500 kpc long inward filament of the E-relic is a peculiar
feature.
It has a width of $\sim 300$ kpc and has steep spectral index.
Backflows from shock fronts, under standard assumptions can have maximum widths
of up to $\sim200-250$ kpc
\citep{bru08}.  The widths of radio relics have been explained as due to such
back flows \citep{bru08,wee10}. However the inward filament in A3376 is
different from such a backflow in that it extends from only 
a section of the E-relic, creating a `notch' at that location. A possibility 
that the `notch' may be due to the breakup of the shock front in an
encounter with a filament has been expressed recently \citep{pau11}. 

A highly polarized filament of extent
$\sim300\times50$ kpc$^{2}$ north of 
the bent-jet radio galaxy with magnetic fields aligned in the same direction
 as in the southern arc of the E-relic is reavealed for the first time. 
This polarized filament is located $\sim200$ kpc behind the
northern part of the E-relic as projected in the sky plane. 

The W-relic, though 
has an arc-like morphology at the edge, lacks an aligned magnetic field.  
It is possible that the turbulence 
in the backflow of the shock has resulted in the complex geometry of the field.
Faraday rotation by the ICM can rotate the polarization vectors. The
relics are located outside the dense part of the ICM (Fig.~\ref{pband}) and thus
this effect is expected to be small \citep{cla04}.
Deeper observations mapping the polarized emission and a model
for the ICM density in A3376 are required.

There are two prominent galaxies in A3376. One is the brightest cluster galaxy
(BCG), 
ESO 307-13,
 and the other is the bent-jet radio galaxy near the peak of the X-ray
emission, MRC $0600-399$ (Fig.~\ref{pband}). Most cD galaxies and BCGs
are located at the cores of galaxy clusters with dense ICM surrounding them
\citep{bau70}. Many of
those are powerful radio galaxies and play a major role in heating of the
ICM. The BCG in A3376 is unique due to its separation of $>900$ kpc from the
peak of X-ray surface brightness.
The powerful cluster-cluster merger in A3376 may have
led to a setting in which the BCG is `orphaned'. Simulation studies may be
able to point out the details of a merger which can result in BCG-ICM peak
separation as large as $>900$ kpc. Mapping of dark-matter distribution in A3376 
with  the gravitational lensing method
is likely to shed additional light on the complex merger dynamics.

A3376 is well suited for a detailed comparison with simulated cluster mergers,
given that it is a well studied cluster at X-rays, optical and radio
wavelengths \citep[][and present work]{aka11b,fli06,bag06}.
Details of the merger
geometry and the properties of individual sub-clusters can be obtained by
matching simulations with observations as has been carried out in the case of
the cluster CIZA J2242.8+5301 \citep{wee11a}. The double relics in CIZA
J2242.8+5301 fit the picture of outgoing merger shocks and 
have been modelled by a toy model of cluster merger shocks.
However, in the case of A3376, due to the morphological details of the E and
W-relics which indicate an interaction with the large scale structure filaments,
a cluster merger in a cosmological simulation will be more appropriate.
This will provide further
insights into the formation of the double relics in A3376 and cluster
mergers in
general.
\subsection{A Radio halo in A3376?}
Radio halos are diffuse radio sources that trace relativistic
electrons and magnetic fields in galaxy cluster centers. Like the relics,
they are also rare and have been found to occur
preferentially in clusters undergoing mergers \citep{cas10}.
It is still not understood why certain merging clusters have radio relics
whereas others have radio halos. Also there are cases where single radio relics 
and radio halos are in the same cluster.
It is important to understand whether the formation of a radio halo 
and of radio relics in a galaxy cluster is simultaneous or it is 
separated in time. This will provide an insight into the details of the 
underlying mechanisms of their generation in cluster mergers. In this
context, we examined whether A3376 has a radio halo. No emission from a  
 radio halo is detected in the 150, 325 and 1400 MHz images.
To obtain upperlimits on the radio power of radio halos a method of injection
of fake radio halos has been used \citep[for e.g.][]{ven08,bru07}. We used AIPS
task `UVSUB' to inject uniform spherical halos of total flux densities in the
range 50 to 500 mJy in the 325 MHz data. The radius of the injected halo was 500
kpc and the location RA 06h01m20.52s DEC -39d59m50.8 (J2000), which is
approximately midway between the two relics. We found that a radio halo of 100
mJy at the level of suspecting diffuse emission (positive residuals) was
detected. Using this as a limiting flux density of detection at 325 MHz, if we
assume a spectral index of -1.0 over the frequency range 325 - 1400 MHz, a limit
on the radio power of 1.02$\times10^{23}$ W Hz$^{-1}$ is obtained at 1400 MHz.
A radio power of 1.91 $\times10^{23}$ W Hz$^{-1}$ at 1.4 GHz is expected if
A3376 were to follow the X-ray luminosity and 1.4 GHz power correlation for
giant radio halos \citep{cas06}. The upperlimit is a factor of $\sim2$ below
the expected radio halo power. Therefore in the picture of radio halo
evolution, this may be either an early stage of halo formation or a fading stage
after attaining the peak radio power. 

\section{Summary}
We have presented a polarization and spectral index study of the double radio
relics in the cluster A3376. The findings from this study can be summarized as 
follows:
\begin{enumerate}
 \item Images of the double relics in A3376 were obtained at 150 and 325 MHz
with the GMRT and at 1400 MHz with the VLA. The integrated spectra of both the
relics show steepening in the frequency range 325 - 1400 MHz as compared to 150
$-$ 325 MHz.
 \item The spectral index distribution (325-1400 MHz) shows flatter spectra at 
the outer edges of both the relics as compared to their inner edges. The Mach
numbers implied by the spectral indices of the relics range between 2.2 - 3.3.
 \item The eastern relic is polarized up to $30\%$ and shows aligned magnetic
field vectors as expected in a shock. The western relic is relatively less
polarized and shows complex orientations of the magnetic fields.
\item The polarization map also reveals a highly polarized filament
located behind the eastern relic as projected in the sky plane, for the first
time. 
\item A steep spectrum filament of size $500$ kpc extends inward from the
eastern relic creating a `notch' at the outer edge of the relic. Such a
structure cannot be explained easily in the standard outgoing shock model for
the double relics. 
\item An upperlimit on the power of a Mpc size radio halo
(log$(P_{1.4GHz}$W Hz$^{-1})<23.0$) is obtained.
\item It is also noted that the brightest galaxy in A3376 is located
$\sim900$ kpc away from the peak of the X-ray emission from the cluster
--possibly `orphaned' by the ongoing merger in the cluster.
\end{enumerate}
 The radio, X-ray and optical
properties of the cluster point toward a major merger in A3376. The
polarization and spectral properties of the relics support the outgoing merger
shock model. A comparison of the properties of A3376 with that of simulated
cluster mergers will lead to a better understanding of the details of the
sub-clusters involved
in the merger. 
\newpage

\section*{Acknowledgments}
We thank the anonymous referee for providing comments that have improved this
paper. We thank the staff of the GMRT who have made these observations
possible. GMRT is run by the National Centre for Radio Astrophysics of the Tata
Institute
of Fundamental Research. The National Radio Astronomy Observatory is a facility
of the National Science Foundation operated under cooperative agreement by
Associated Universities, Inc. This research has made use of the NASA/IPAC
Extragalactic Database (NED) which is operated by the Jet Propulsion Laboratory,
California Institute of Technology, under contract with the National Aeronautics
and Space Administration. We have made use of the ROSAT Data Archive of the
Max-Planck-Institut fur extraterrestrische Physik (MPE) at Garching, Germany.

\bibliographystyle{mn}
\bibliography{a3376}

\bsp

\label{lastpage}

\end{document}